\documentclass{aastex}
\usepackage{spr-astr-addons}             
\usepackage{url}
\usepackage{lscape, morefloats, graphicx, float, longtable}
\usepackage{graphicx}
\RequirePackage{color}                   

\begin{document}

\title{Local Stellar Kinematics from RAVE data -- VIII. Effects of the Galactic Disc Perturbations on Stellar Orbits of Red Clump Stars}
\slugcomment{Not to appear in Nonlearned J., 45.}
\shorttitle{Local stellar Kinematics from RAVE data - VIII.}
\shortauthors{\"O. \"Onal Ta\c s, S. Bilir, O. Plevne}

\author{\"O. \"Onal Ta\c s \altaffilmark{1}}
\altaffiltext{1}{Istanbul University, Faculty of Science, Department 
of Astronomy and Space Sciences, 34119 University, Istanbul, Turkey\\
\email{ozgecan.onal@istanbul.edu.tr}}

\author{S. Bilir \altaffilmark{1}}
\altaffiltext{1}{Istanbul University, Faculty of Science, Department 
of Astronomy and Space Sciences, 34119 University, Istanbul, Turkey\\}

\and
\author{O. Plevne \altaffilmark{2}}
\altaffiltext{2}{Istanbul University, Department of Astronomy and Space 
Sciences, Graduate School of Science and Engineering, Istanbul University, 
34116, Beyaz{\i}t, Istanbul, Turkey\\}

\begin{abstract}  
We aim to probe the dynamic structure of the extended Solar neighbourhood by calculating the radial metallicity gradients from orbit properties, which are obtained for axisymmetric and non-axisymmetric potential models, of red clump (RC) stars selected from the RAdial Velocity Experiment's Fourth Data Release. Distances are obtained by assuming a single absolute magnitude value in near-infrared, i.e. $M_{Ks}=-1.54\pm0.04$ mag, for each RC star. Stellar orbit parameters are calculated by using the potential functions: (i) for {\it the MWPotential2014} potential, (ii) for the same potential with perturbation functions of the Galactic bar and transient spiral arms. The stellar age is calculated with a method based on Bayesian statistics. The radial metallicity gradients are evaluated based on the maximum vertical distance ($z_{max}$) from the Galactic plane and the planar eccentricity ($e_p$) of RC stars for both of the potential models. The largest radial metallicity gradient in the $0<z_{max} \leq0.5$ kpc distance interval is $-0.065\pm0.005$ dex kpc$^{-1}$ for a subsample with $e_p\leq0.1$, while the lowest value is $-0.014\pm0.006$ dex kpc$^{-1}$ for the subsample with $e_p\leq0.5$. We find that at $z_{max}>1$ kpc, the radial metallicity gradients have zero or positive values and they do not depend on $e_p$ subsamples. There is a large radial metallicity gradient for thin disc, but no radial gradient found for thick disc. Moreover, the largest radial metallicity gradients are obtained where the outer Lindblad resonance region is effective. We claim that this apparent change in radial metallicity gradients in the thin disc is a result of orbital perturbation originating from the existing resonance regions. 
\end{abstract}


\section{Introduction}
Galaxies are dynamically active ensembles of gas, dust and stars that take aeons to shape their almost stationary faces. Likewise our galaxy, the Milky Way, is exposed to certain drastic changes due to both inner and outer perturbations. The Galaxy is generally considered to have distinct components, which are known as bulge, disc and halo, and each have different characteristics in age, kinematics and chemistry. Regardless of the status of the transition between these Galactic components, they are formed from each other as a combination of gas either in clouds or in dark matter halos inside the same gravitational potential \citep{BHG16}. These components are thought to be shaped by an on-going collapse into a plane of the gas material, or by part of stellar evolution of its subjects and or by merger events with the surrounding environment such as dwarf galaxies, gas etc \citep{BT08}. Meanwhile, the gas content governs this formation period in two ways. Firstly by its amount and how it varied over time and secondly by the chemical enrichment of the gas via on-going stellar evolution of the stars \citep*{CMG97}. It is known that stars have the same chemical structure as the molecular cloud in which they were born until the dredge-up and/or stellar wind processes take over. Moreover, as a natural extension of the stellar evolution, most of the synthesized elements do return to the interstellar medium (ISM), where new stars are formed, either through supernova explosions of short-lived massive stars and/or binary stars or expanding shells of long-lived low mass stars. These processes enrich the ISM with various elements in the periodic table on different time-scales and abundances. This is known with different names such as chemical tagging \citep{Freeman02}, the cosmic clock approach \citep{MCM13}. This, in principle, helps to uncover the time dependent chemical structure of any chosen, observable Galactic region, and also helps to put constraints on Galactic formation scenarios. At this juncture, one other important aspect of the Galactic evolution helps to establish the necessary environment to test the existing findings regarding the Galactic formation, which is called as the inside-out formation of the Galaxy disc \citep[][and references therein] {CMG97}. According to the two infall model \citep{CMG97, Chiappini01} the thin disc is evolved  by the infall of high-angular momentum gas from inside-out, from the central regions like bulge to the outer regions of the disc radius, for the last 7-8 Gyrs.

The Sun is embedded in the Galactic disc and since the disc entails two important aspects of the chemo-dynamic evolution; on one hand it is composed of 80 per cent of the visible matter in the Galaxy and on the other it contains the major perturbers of the Galactic potential, such as the Galactic bar and spiral arms, therefore it is only logical to further study it. Over the past two decades there has been various type of all sky surveys, i.e. {\it Hipparcos} \citep{ESA97}, SDSS \citep{York00}, GCS \citep{Nordstrom04}, 2MASS \citep{Skrutskie06}, RAVE \citep{Steinmetz06}, APOGEE \citep{Allende08}, SEGUE \citep{Yanny09}, GALAH \citep{deSilva15}, LAMOST \citep{Zhao12}, GES \citep{Gilmore12}, {\em Gaia} \citep{Gaia16}, which swept various sized space volumes around the Sun and gathered a huge amount of astrometric, photometric and spectroscopic data of the objects in it. Meanwhile, our increasing capability in theoretical computation allowed us to develop models in order to test the partial or complete formation and evolution scenarios of the Milky Way and its components in different contexts \citep[cf.][]{Minchev17, Bovy15, Hayden15, Nidever14, MCM13}.

Observations of the Solar neighbourhood objects allowed us to spot the various effects of these interactions by studying their chemistry, kinematics and orbital dynamics. According to \citet{ELS62}, stellar orbit shapes vary from circular to elongated eccentric as the vertical distance from the Galactic plane increase. \cite{Wielen77} showed that the natural encounters with massive objects like giant molecular clouds during a lifetime of a star do alter the shape of the stellar orbit. Moreover, the Galactic bar causes resonance regions that affect the gravitational field which are known as inner and outer Lindblad (ILR and OLR) and co-rotation resonance (CR) regions. In CR region stars revolve with the Galactic bar with the same angular rotational rate, while the interplay between azimuthal and radial angular frequencies define whether a star is in the ILR or OLR \citep [cf.][]{Dehnen00, Minchev17}. According to \cite{Fux99}, the CR region covers 3.5 - 5 kpc of Galactocentric distance range. Based on the hydrodynamical modelling of the inner Galaxy, \citet{Dehnen00} suggests that the Solar neighbourhood is in the OLR of the Galactic bar. \citet{Dehnen00} showed that under the assumption of a flat rotation curve $R_{OLR}=1.7 R_{CR}$ and using the epicycle approximation for late-type disc stars roughly calculated the Galactocentric position of the OLR as 6 - 9 kpc. Also, various values for OLR's position is quoted in \citet{Antoja09, Antoja15}'s stellar stream studies.

Metallicity gradients are one way of looking into the chemo-dynamic structure of the Galaxy, which is a stronger sight that  probe the change in metallicity trends during its ongoing evolution, especially if various metallicity indicators are derived from the mid- or high-resolution spectroscopy (HRS) of a large number of objects that reach out to a large space volume. Yet it is very hard to reveal the complexities that govern the temporal evolution of the Milky Way, only by looking into the metallicity gradients. It is important to have accurate spectroscopic observations and distance measurements for metallicity gradient calculations. Even though the metallicities become more and more reliable with each new generation of the multi-object spectroscopic surveys, the distances still remain problematic. There are limited number of HRS studies such as APOGEE, GALAH, GES in the literature and these surveys do not cover all sky, thus stars with accurate metallicities are only limited to a few kpc within the Solar neighbourhood. Especially, these limitations decrease the sensitivity of the radial metallicity gradient calculations. Apart from the common interest in the current Galactocentric distances, analyze from the current position of stars do not exactly reflect their true Galactic population whereabouts. To overcome this problem there is another approach in the literature, in which the Galactic orbit parameters of various objects are calculated, and by doing this the sensitivity of the radial metallicity gradients is enhanced. Even though in recent studies \citep[i.e.][]{Nordstrom04, Wu09, Coskunoglu12, Bilir12, Boeche13, Boeche14, Plevne15, OnalTas16} that determine the radial metallicity gradients using this method, the gradient values still scatter in a wide range. This variation might depend on the spectral resolution, object type, distance determination, and Galactic populations. On the other hand, the most crucial dependencies stem from the spectral analysis method, such as spectral fitting, equivalent widths, atmospheric models, local or non-local thermodynamic equilibrium. The majority of the metallicity gradient studies between the years 2000 and 2016 are compiled in \cite{OnalTas16}. Also, in the same study the authors pointed out that the radial metallicity gradients are strongly dependent on both the maximum vertical distances from the Galactic plane and the planar eccentricities. Moreover, stellar orbits are affected by Galactic perturbation sources with age, and this makes the orbits more eccentric, which in result changes the Galactic radial metallicity gradients in time.

This paper is aimed to show the effects of major Galactic perturbers, such as the Galactic bar and spiral arms, on stellar orbits of the RC stars selected from the RAdial Velocity Experiment's Fourth Data Release \citep[RAVE DR4;][]{Kordopatis13}. It is also shown that the planar eccentricities play a crucial role in selecting the proper sample in order to reflect the chemo-dynamic structure of the Solar neighbourhood. The layout of the paper is as follows. In Sections 2 and 3, the data and methods are introduced, in Section 4, the results are presented and finally, the results are discussed in Section 5.

\section{Data}
RC stars are selected from the RAVE DR4 catalogue \citep{Kordopatis13}. We neglected stars with following properties from the catalogue: (i) stars with missing model atmosphere parameters, (ii) stars with missing proper motion components from the fourth U.S. Naval Observatory CCD Astrograph Catalog \citep[UCAC4,][]{Zacharias13}, (iii) stars which do not have 2MASS photometric filter quality as $A$, which also means signal-to-noise ratio is ${\it S/N}<20$ \citep{Cutri03}, and (iv) stars that observed in Galactic latitude $|b|<10^{\circ}$, to avoid the effects of the interstellar medium, (v) stars with radial velocity errors less than 10 km s$^{-1}$, (vi) stars with \textsc{algo\_conv=1}, which means spectral pipeline has not converged, and (vii) stars that do not have ``normal'' spectral morphology. After these mandatory eliminations, the initial selection criteria is applied on the model atmosphere parameters of the remaining sources in the RAVE DR4 catalogue \citep{Kordopatis13}, i.e. $4000<T_{eff}\,{\rm (K)}<5400$ and $1<\log~g\,{\rm (cgs)}<3$. In Fig. 1, Hertzsprung - Russell (HR) diagrams of the region where the RC stars are most likely to reside are presented in two panels, one colour coded by logarithmic number density and the other by logarithmic metallicity. Also, a 1$\sigma$ contour line based on the most crowded region is drawn with white solid line in Fig. 1. Using the contour line, we selected the stars that reside in the 1$\sigma$ area. Moreover, we applied ${\it S/N}\geq 40$ constraint, which reduced the number of sources to 51,941.

\begin{figure}[t]
\centering
\includegraphics[width=8cm, height=6cm]{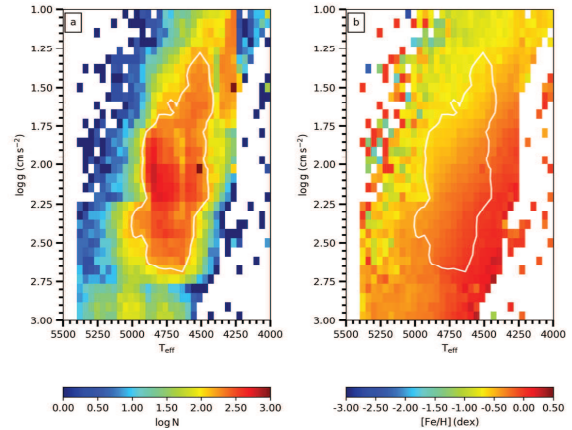}
\caption{HR diagram of selected RC stars from RAVE DR4, colour-coded by logarithmic number density (a) and by metallicity (b).} 
\label{Fig01}
\end{figure} 

RC stars are considered as standard candles for the distance determination, thus their absolute magnitudes are assumed to be constant, especially in near-infrared $K_s$ band, $M_{Ks}=-1.54\pm0.04$ mag \citep{Groenewegen08}. Colour excess at infinity values, $E_{\infty}(B-V)$, for each star are obtained individually from \cite{SF11}'s reddening maps from NASA IPAC Extragalactic database and are reduced to $E_{d}(B-V)$ related to individual stellar distances of the RC sample using \citet{Bahcall80}'s equation. Finally, the distance and $E_{d}(B-V)$ values are calculated using the photometric parallax method. This method is described in detail in the studies of \citet{Coskunoglu12} and \citet{Bilir12}.

\section{METHODS}

\subsection{Stellar Kinematics}

When RAVE DR4's radial velocities ($\gamma$), UCAC4 proper motions ($\mu_{\alpha} \cos \delta$, $\mu_{\delta}$) and distances ($d$) are coupled with individual equatorial and Galactic coordinates of the RC sample in accordance with the algorithm and matrices of \cite{Johnson87}, then the space velocity components ($U$, $V$, $W$) and their respective uncertainties are calculated for the J2000 epoch. In a right-handed Galactic reference system, the space velocity components are defined as $U$ positive towards the Galactic centre, $V$ in the direction of Galactic rotation, and $W$ positive towards the North Galactic Pole. The total space velocity and its uncertainty are calculated as root mean squares of the individual space velocity components and their uncertainties, respectively, for each RC star. Apparent magnitudes of RAVE stars vary in the $9<I<12$ mag interval and the RAVE catalogue includes stars with different luminosity classes, especially giants and dwarfs. When their distances are compared, one see that giant stars are dispersed in a larger space volume than dwarf stars \citep{Plevne15, OnalTas16}. Thus, it is important to make a correction on the space velocity components that are parallel to the Galactic plane for the differential rotation, especially for the bright objects in the sample. Differential rotation corrections of \citet{Mihalas81} are applied only on the $U$ and $V$ space velocity components in order to compensate the distance divergence on the orbital speeds around the Galactic centre. However, the $W$ space velocity component is not corrected for differential rotation since it is aligned vertically to the Galactic plane. Then all of the space velocity components are reduced into the local standard of rest of \citet{Coskunoglu11}'s, $(U_{\odot}, V_{\odot}, W_{\odot})_{LSR}=(8.83\pm 0.24, 14.19\pm 0.34, 6.57\pm 0.21)$ km s$^{-1}$. 

Uncertainties in each space velocity component are calculated by propagating uncertainties in radial velocity, distance and proper motion of RC stars using \cite{Johnson87}'s algorithm. The total space velocity errors are calculated as the square root of sum of the squares of the individual space-velocity component errors, i.e. ($S_{err}=\sqrt{U^2_{err}+V^2_{err}+W^2_{err}}$). The total space velocity error distribution is shown in Fig. 2a. A final constraint is applied on the total space velocity errors by defining a cutoff point. To do this, the median and 1$\sigma$ values of the $S_{err}$ distribution, 9.42 and 10.54 km s$^{-1}$, respectively, are added to obtain the cut-off point, which give $S_{err}\simeq 20$ km s$^{-1}$ as a limiting value. Thus, stars with total space velocity errors greater than 20 km s$^{-1}$ are removed from the sample, which is only 10\% of 51,941 RC stars. So, we retained the most sensitive 46,818 RC star sample. After these eliminations final RC sample has median space velocity errors: ($U_{err}, V_{err}, W_{err}=(4.89\pm 2.67, 4.09\pm 2.87, 4.88\pm 2.75)$ km s$^{-1}$. The histograms of the space velocity component errors are shown in Fig. 2b-d. Finally,  histograms of the model atmosphere parameters ($T_{eff}$, $\log g$, [Fe/H]) of the RC sample after the imposing on the total space velocity error are presented in the three panels of Fig. 3.

\begin{figure}
\centering
\includegraphics[width=8cm, height=15cm]{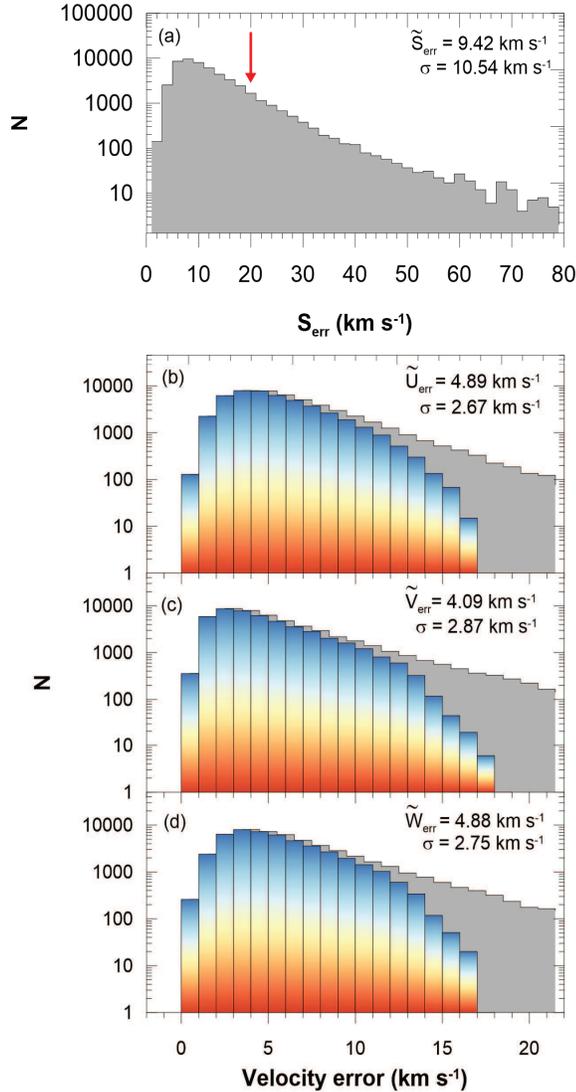}
\caption{Total space velocity error histogram of 51,941 stars ({\it top panel}). The median and standard deviation are 9.42 km s$^{-1}$ and 10.54 km s$^{-1}$, respectively. Their sum gives approximately 20 km s$^{-1}$ (red arrow) and this is the last constraint that imposed on the sample. As a result 46,818 RC stars are retained as the final sample. Histograms of $U$, $V$, and $W$ space velocity errors are shown in comparison with the total space velocity error histogram of 46,818 RC stars ({\it lower panels}).}
\label{Fig06}
\end {figure} 

\begin{figure}[h]
\centering
\includegraphics[scale=0.40, angle=0]{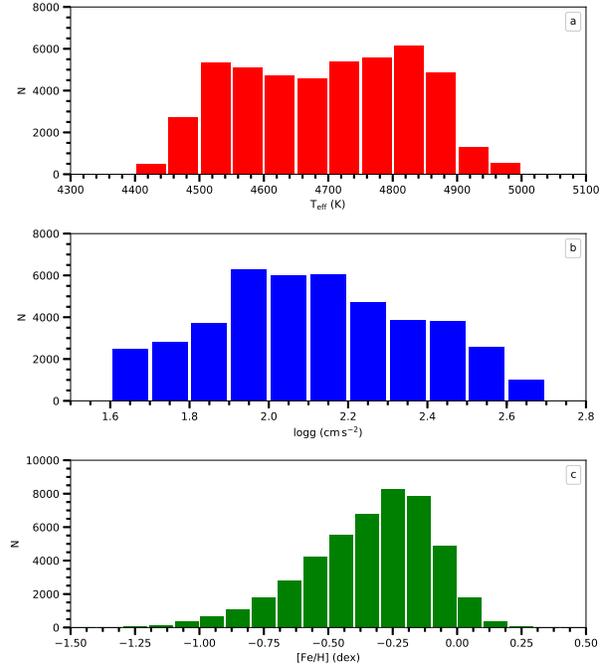}
\caption{Histograms of the model atmosphere  parameters of 46,818 RC stars in RAVE DR4. Effective temperature (a), logarithmic surface gravity (b), and  metallicity (c).} 
\end{figure} 

\subsection{Stellar Dynamics}
Properties of RC orbits are calculated using the potential functions that defined in {\it galpy}, the Galactic dynamics library \citep{Bovy15}. These calculations are performed both for an axisymmetric potential for the Milky Way galaxy (hereafter MW) and for a non- axisymmetric potential, which includes the perturbation functions of the Galactic bar and the spiral arms (hereafter MWBS). Galactic rotational speed and the Galactocentric distance of the Sun are chosen as 220 km s$^{-1}$ and 8 kpc \citep{Majewski93}, respectively, for the dynamic parameter calculations.

The MW potential is selected from the {\it galpy.potential} library, which has the relevant parameters of the Milky Way galaxy, also known as {\it MWPotential2014}, \cite[for details see Table 1 in][]{Bovy15}.  This potential function is composed of three models that each of which represents the gravitational field of one of the main Galactic components of the Galaxy, i.e. bulge, disc, and halo. According to \citet{Bovy15}, the bulge is modeled as a spherical power law  density profile with exponential cut off at 1.9 kpc with an exponent of -1.8, which is shown in Eq. (1),

\begin{equation}
\rho (r) = A \left( \frac{r_1}{r} \right) ^{a} \exp \left[-\left(\frac{r}{r_c}\right)^2 \right] \label{eq:rho}
\end{equation} 
where $A$, is the amplitude that applied to the potential in mass density units, $\alpha$ is the inner power, $r_c$ and $r_1$ are the cut-off radius and the reference radius for the amplitude, respectively.

The disc is modeled with a \citet{MN75} potential as in Eq. (2),
\begin{equation}
\Phi_{disc} (R_{gc}, z) = - \frac{G M_d}{\sqrt{R_{gc}^2 + \left(a_d + \sqrt{z^2 + b_{d}^2 } \right)^2}} \label{eq:disc}
\end{equation}
here $R_{gc}$, is star's distance from the Galactic centre, $z$ is star's vertical distance to the Galactic plane, $G$ is the Universal gravitational constant, $M_d$, is the disc mass, $a_d$ and $b_d$ are the scale-length and the scale-height of the Galactic disc, respectively. 

The Galactic halo is modeled with \citet*{NFW96} potential defined as in Eq. (3),

\begin{equation}
\Phi _{halo} (r) = - \frac{G M_s }{R_{gc}} \ln \left(1+\frac{R_{gc}}{r_s}\right) \label{eq:halo}
\end{equation} 
where $M_S$ and $r_s$ are the mass and the radius that assumed for the dark matter halo of the Galaxy.

The perturbation function of the Galactic bar is defined with \citet{Dehnen00}'s bar potential as in Eq. (4),

\begin{align}
\Phi (R, \phi) = A_b(t) \cos (2(\phi - \Omega_b t)) \\ \nonumber
\times 
\begin{cases}
  -(R_b/R)^3 \, &  {\rm for} \, R > R_b   \\
  (R_b/R)^3 -2 \, & {\rm for} \, R \leq R_b
	\end{cases}
\label{eq:cases}
\end{align}
here $A_b(t)$, $\Omega_b$ and $R_b$ are the amplitude, the rotation speed and the semi-major axis of the Galactic bar and $\phi$ is the angle between the semi-major axis and the line from Sun to the Galactic centre. The perturbation function of the spiral arms are selected as transient logarithmic spiral arm potential from the {\it galpy} library, which is shown in Eq. (5),

\begin{equation}
\Phi (R, \phi) = \frac{amp(t)}{\alpha} \cos ( \alpha \ln R - m (\phi - \Omega_s t-\gamma)) \label{eq:phi}
\end{equation}
where the term $amp(t)$ is the time dependent amplitude of the potential and modeled by a Gaussian $amp(t)=amp \times  A \, \exp(-(t-t_o)^2/(2\sigma)^2)$ in which $t_o$ is the time at which the spiral peaks and $\sigma$ is the spiral duration. Other parameters, such as $\Omega_s$, $\alpha$, $m$ and $\gamma$ are the pattern speed, the pitch angle, the number of spiral arms and the angle between the line connecting the peak of the spiral pattern at the Solar radius and the line from Sun to the Galactic centre, respectively.

Each star's orbital properties around the Galactic centre are extracted from the complete orbits that are calculated with 3 Myr steps for a 3 Gyr integration time. Same input parameters that are used in space velocity component calculations are also used in dynamic orbit parameter estimations. According to this, the output parameters are peri- and apo-galactic distances ($R_p$ and $R_a$), the planar eccentricity ($e_p$) and the maximum vertical distance from the Galactic plane ($z_{max}$). Moreover, the mean Galactocentric distance ($R_m$) is calculated as an average of $R_p$ and $R_a$. These are the key parameters for investigating the effects of the large-scale Galactic perturbation forces, i.e. bar and spiral arms, in this study. When propagated errors, that are caused by radial velocities, proper motions and distances are taken into account, we found the mean errors on $z_{max}$, $R_m$, and $e_p$ of RC stars to be $0.1$ kpc, $0.2$ kpc, and $0.02$, respectively.

Galactic orbit parameters of 46,818 RC stars are calculated for the MW and MWBS potential models and their comparison is presented in Fig. 4. Mean differences and their standard deviations for peri- and apo-galactic distance comparisons for both potential models are 0.16, 0.37 kpc and 0.22, 1.12 kpc, respectively. Similarly, the mean difference and it's standard deviation for the mean Galactocentric distances are found from comparison as 0.17 and 0.55 kpc. Moreover, the same calculation is performed for the planar eccentricity, which is a function of $R_p$ and $R_a$, and it's mean difference and standard deviation are 0.033 and 0.043. In Fig. 4c, $R_m$ values of RC stars are compared. There are obvious horizontal strips in different Galactocentric distances. The stepped structure consisting of these horizontal strips is formed by RC stars caught in Lindblad resonances at different distances. Moreover, in Fig. 4d, in which compares the planar eccentricities, there is a separate population of RC stars that reside between $0.3<e_p<0.6$ in MW axis and $0.6<e_p<0.8$ in MWBS axis. This prominent RC group is also apparent in the lower left part of the Fig. 4b, which shows the comparison between the perigalactic distances. When the maximum vertical distance from the Galactic plane and the metallicity abundances of this group of stars are examined, we found that they lie between $0.8<z_{max}<10$ kpc distance interval and their mean distance is $\langle z_{max}\rangle=1.22\pm0.60$ kpc and their metallicities range is $-1.0<[{\rm Fe/H}]<-0.3$ dex and their mean metallicity is $\langle[{\rm Fe/H}]\rangle=-0.66\pm0.30$ dex. These results point out that this group of stars belong to other stellar populations in the Galaxy, i.e. mostly thick disc and halo. Since population analysis did not applied on the main RC sample, this is not a very surprising result and it can be expected to see small number of stars from different stellar populations. There are 382 out 46,818 RC stars with properties that deviate from the general trends of the main sample. Moreover, this small number of stars is less than 1\% of the total sample, thus they do not directly affect the results obtained from the analysis. It is striking that there are significant differences and large standard deviations among Galactic orbit parameters calculated from axisymmetric and non-axisymmetric Galactic potentials and this lead us to evaluate the results obtained from both of the potentials individually.

\begin{figure}[h]
\centering
\includegraphics[width=8cm, height=8cm]{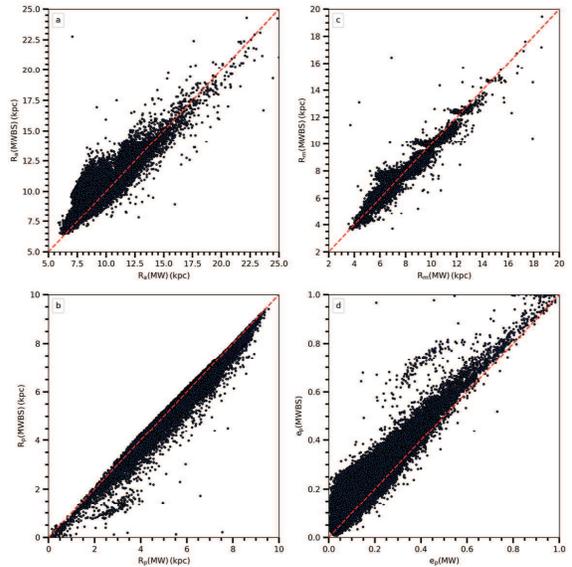}
\caption{Comparison between Galactic orbital parameters calculated using the MW and MWBS potential models. Apo-galactic distance (a), peri-galactic distance (b), mean Galactocentric distance (c), and planar eccentricity (d).}
\label{Fig04}
\end{figure} 

\subsection{Stellar Ages}

Stellar positions, space velocities, chemical structure and ages are necessary parameters to thoroughly investigate the existing structure and the evolution of the Milky Way  \citep[cf.][]{Freeman02, BHG16}. In this study, RC ages are calculated with a method based on Bayesian statistics, which includes probability density functions (pdf) that are obtained from comparative calculations of the theoretical model parameters and the observational parameters \citep{PE04, JL05, Duran13}. According to \citet{PE04}, the Solar neighbourhood mass, age and metallicity priors are assumed to be independent, therefore they are considered separately. Based on this argument, the mass distribution of RC stars is important. In our RC sample, stars reside in a narrow mass range \citep[$0.8\leq M/M_{\odot}\leq2.2$;][]{Girardi99}, so the choice of the IMF has no strong influence on Bayesian age calculations. In the recent study of \citet{Girardi16}, the IMF for a typical RC star in the Solar neighbourhood is 1.3 $M_{\odot}$. Since the SFR of the Galactic disc is not well known, it is assumed to be either constant or slowly decreasing \citep[cf.][]{PE04}. Moreover, the reliability of the metallicity prior in age calculations for the Solar neighbourhood sample depends on the uncertainties. Our sample has $\sigma_{[{\rm Fe/H}]}\simeq 0.10$ dex \citep{Kordopatis13}, thus the metallicity affects age calculations of RC stars less, since the uncertainties are rather small. Likewise following \citet{JL05}, the first assumption was that the standard errors in the observational parameters are independent of each other and show normal distributions. Then a likelihood function (${\cal L}$) is defined as

\begin{equation}
{\cal L}(\tau, \zeta, m) = \prod_{i=1} ^{n} \frac{1}{\sqrt{2 \pi} \sigma_{i}} \exp{[-\chi ^2 /2]}
\label{eq:likelihood}
\end{equation}
where 
\begin{equation}
\chi ^2 = \sum_{i=1}^{n} \left( \frac{q_{i}^{obs} - q_{i}(\tau , \zeta , m) }{ \sigma_i} \right)
\label{eq:chi2}
\end{equation}
here $q$ represents the model atmosphere parameters, such as the effective temperature ($T_{eff}$), the logarithmic surface gravity ($\log g$) and the metallicity ([Fe/H]), whereas $\tau$, $\zeta$ and $m$ are the theoretical model parameters which represents age, metallicity, and mass, respectively. $n$ and $\sigma$ stand for the number of objects and the standard errors, respectively. Then, the final pdf is obtained by applying a Bayesian correction defined as
\begin{equation}
f (\tau , \zeta , m) \varpropto f_0  (\tau, \zeta, m) \times {\cal L}(\tau, \zeta, m)
\label{eq:bayesian}
\end{equation}
where ${f_0}$ is the initial pdf and given as
\begin{equation}
f_0=\psi(\tau) \varphi (\zeta | \tau) \xi(m | \zeta , \tau)
\label{eq:priors}
\end{equation}
here $\psi (\tau)$, $\varphi (\zeta)$ and $\xi (m)$ are the star formation rate (SFR), the metallicity distribution (MDF) and the initial mass function (IMF), respectively. SFR and MDF are assumed to be constant since $\tau$, $\zeta$ and $m$ are independent from each other. However, the IMF is expressed as a power law, \citep{JL05}, because the amount of low-mass stars is bigger than the high-mass stars in the Galaxy. The final pdf in Eq. (8) is also depending on other parameters along with age and this is overcome by integrating the final pdf over the MDF and the IMF. When Eq. (8) is inserted in Eq. (9) and integrated, one can obtain the $G(\tau)$ function which shows the pdf obtained from a comparison between observationally inferred model atmosphere parameters and theoretical model parameters.

\begin{equation}
G(\tau) \varpropto \int \int {\cal L}(\tau, \zeta, m) \xi (m) dm d\zeta
\label{eq:Gfunction}
\end{equation}

According to \citet{JL05}, Eq. (10) is evaluated for each age value ($\tau_j$) as a double sum along a set of isochrones at the required age that are equidistant in metallicity ($\zeta_k $). In our application, we used pre-computed isochrones for a step size 0.05 dex in $\zeta$. Let $m_{jkl}$ be the initial mass values along each isochrone ($\tau_j, \zeta_k $). This function is re-formed by applying isochrone data into the function and comparing the results with observational parameters given in the below form,

\begin{equation}
G(\tau_j) \varpropto  \Sigma_k \Sigma_l {\cal L}(\tau_j , \zeta_k , m_{jkl}) \xi (m_{jkl}) (m_{jkl+1} - m_{jkl-1})
\label{eq:Gfuntion_numerical}
\end{equation}

The function $G$ is obtained for a large age range by comparing model atmosphere parameters with each of the isochrones, but the maximum of $G$ function gives the most likely age for any given object. Isochrones are selected from the PARSEC \citep{Bressan12} stellar evolution models for $-2.25<$[Fe/H](dex)$<+0.5$ and $0 <\tau \, {\rm (Gyr)}<13$ with 0.05 dex and 0.1 Gyr steps, respectively. The results of Bayesian age estimations of 46,818 RC stars are shown in Fig. 5. In the figure the frequency counts of the age parameter is shown in five metallicity intervals, namely, $0<$[Fe/H]$\leq+0.5$, $-0.25<$[Fe/H]$\leq0$, $-0.5<$[Fe/H]$\leq-0.25$, $-0.75<$[Fe/H]$\leq-0.5$, and $-1.5<$[Fe/H]$\leq-0.75$ dex, from the top panel to the bottom, respectively.

In the RAVE DR4 catalogue \citep{Kordopatis13}, the internal errors for the atmosphere model parameters are given in tabular form after re-normalization with photometric priors as a function of spectral type and luminosity class. Internal errors of atmosphere model parameters for KII-V and $S/N>50$, which are listed in Table 2 of \citet{Kordopatis13}, are adopted general errors for individual star in our RC sample, since they are mostly KIII stars with $S/N>40$, i.e. $76\leq T_{eff} (\rm K)\leq 105$, $0.14\leq \log g (\rm cgs)\leq 0.35$, $0.08\leq \rm [M/H](dex)\leq 0.10$. Due to complex and non-Gausian nature of Bayesian age estimation method, overall uncertainties on estimated RC star ages vary between 20\% and 50\%.

\begin{figure}[h]
\centering
\includegraphics[scale=0.37, angle=0]{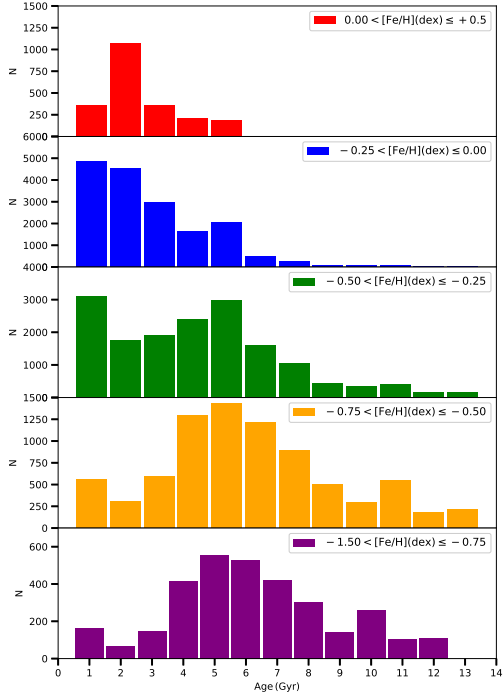}
\caption{Stellar age distribution of RC stars in the five metallicity intervals: $0<$[Fe/H]$\leq+0.5$, $-0.25<$[Fe/H]$\leq0$, $-0.5<$[Fe/H]$\leq-0.25$, $-0.75<$[Fe/H]$\leq-0.5$, and $-1.5<$[Fe/H]$\leq-0.75$ dex from top to bottom panel, respectively.} 
\end{figure} 

\section{Results}

Various parameters of RC stars in RAVE DR4 such as distances, space velocities, orbit properties and ages are calculated with the methods described above in order to assess the major perturbation effects on the stellar orbits that govern the Galactic disc. The RC stars are not separated into stellar populations with any existing methods in the literature, but instead they are roughly separated based on their maximum vertical distance from the Galactic plane. In this section, the results obtained for the radial metallicity gradients calculated both for the MW and MWBS potentials are presented.

\subsection{Radial Metallicity Gradients}

The radial metallicity gradients of the RC stars are calculated for the mean Galactocentric distances obtained from their completed orbits in 3 Gyr around the Galactic centre, whereas the metallicity parameters are obtained from RAVE DR4. In this study, the radial metallicity gradients are investigated as a function of $z_{max}$ and $e_p$. Most of the RC sample ($\sim$99\%) reaches up to $z_{max}=3$ kpc from the Galactic plane, thus the RC stars are divided into four consecutive $z_{max}$ distance intervals, for the MW and MWBS potential models, each of which reflects a certain state of the Galactic disc. $z_{max}$ distance limits are defined by considering the population efficiencies, so the RC sample is separated into following distance intervals: $0<z_{max}\leq0.5$ kpc for dominant thin-disc stars, $0.5<z_{max}\leq 1$ kpc for the transition region between thin and thick disc stars, $1<z_{max}\leq 2$ kpc for dominant thick-disc stars and $z_{max}>2$ kpc for the transition region between thick disc and halo stars. Then, the $e_p$ distribution in these samples are investigated with two different approaches: First, we grouped RC stars with cumulatively increasing $e_p$ subsamples, which allows us to see the contamination effects on the radial metallicity gradients from the inclusion of the eccentric stellar orbits and second, we considered discretely increasing $e_p$ subsamples, which gives the radial extent of the RC samples that are affected from the well-known resonance regions (CR and OLR). In both approaches each $e_p$ subsample is selected between 0 and 1 with $\Delta{e_p}=0.05$ steps. The cumulative $e_p$ distributions in each $z_{max}$ distance interval for the MW and MWBS potential models are shown in the left and right panels in Fig. 6, respectively. According to both panels, the number of stars reaches up to 90\% at $e_p=0.5$, for $z_{max}\leq2$ kpc distances, while for $z_{max}>2$ kpc this happens at $e_p=0.6$. However, the $e_p$ subsamples beyond the $e_p=0.5$ limit are excluded in all $z_{max}$ distance intervals. By applying this limitation on the RC sample, 2\% and 4\% of stars are removed from the sample for MW and MWBS models, respectively. 

\begin{figure*}
\centering
\includegraphics[trim=2cm 0cm 1cm 1cm, clip=true, scale=0.30, angle=0]{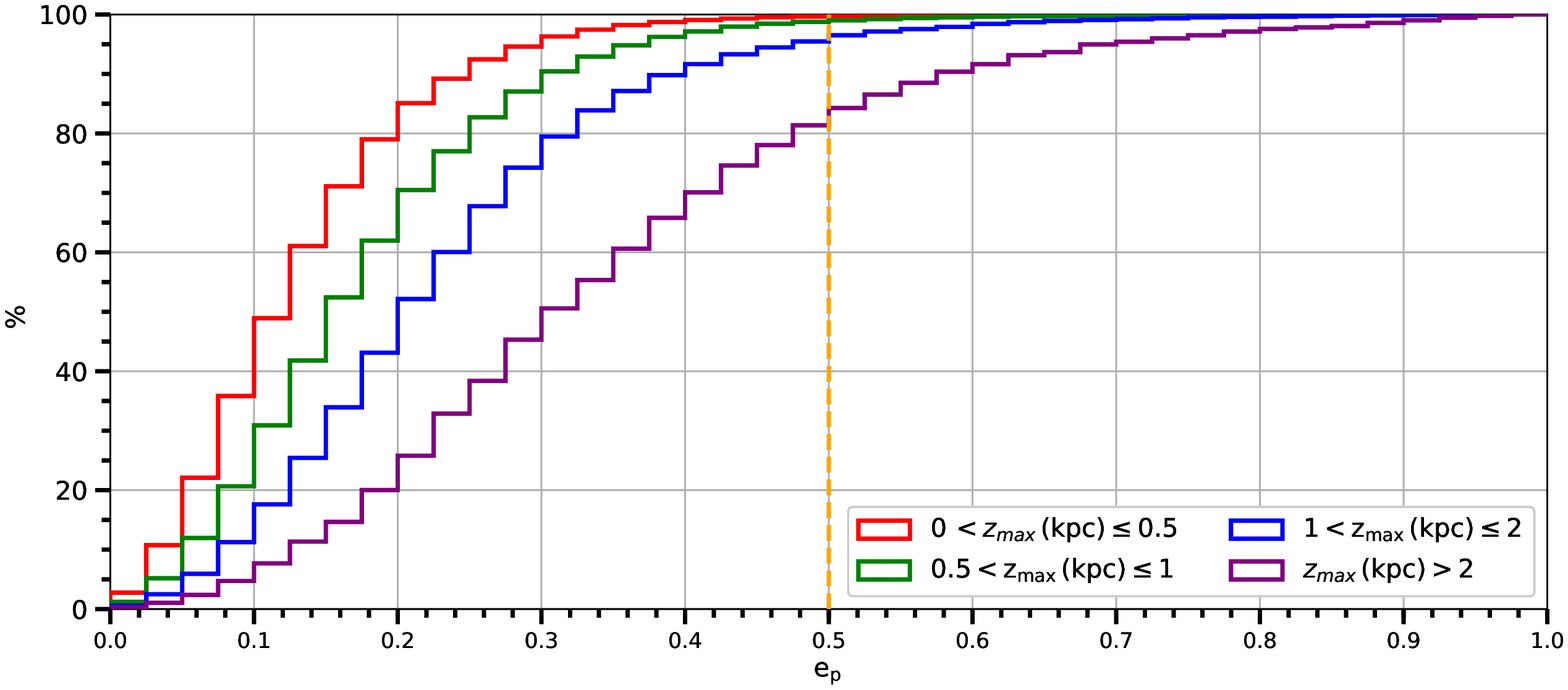}
\includegraphics[trim=2cm 0cm 1cm 1cm, clip=true, scale=0.30, angle=0]{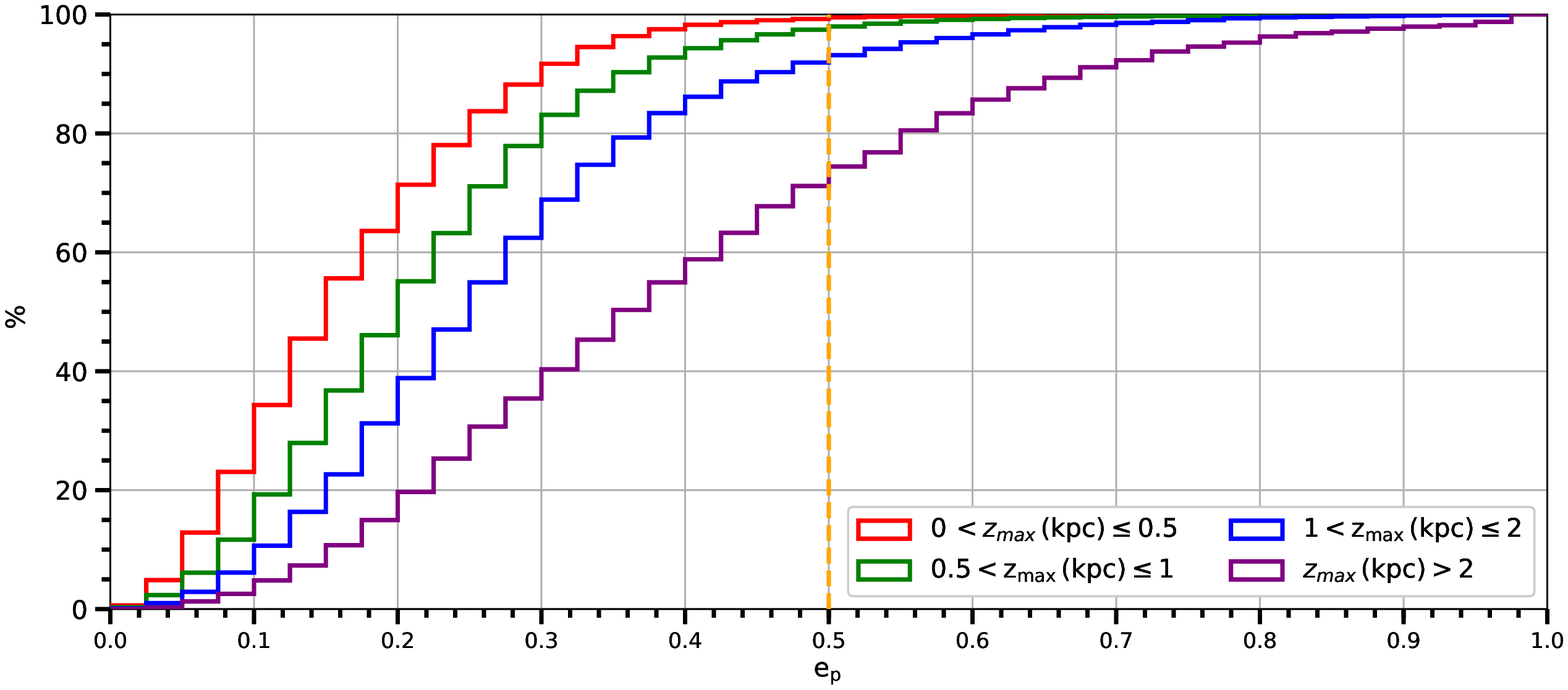}
\caption{Cumulative number of RC stars in the $e_p$ parameter for the MW (left panel) and MWBS (right panel) potential models. The solid lines represent $0<z_{max}\leq0.5$ (red), $0.5<z_{max}\leq1$ (green), $1<z_{max}\leq 2$ (blue) and $z_{max}>2$ kpc (purple) distance intervals. The orange dashed lines represent the $e_p=0.5$ limit where at least 90\% of the RC stars reside in the first three distance intervals.} 
\end{figure*} 

\subsubsection{According to MW Potential Model}

The radial metallicity gradients which are calculated for the RC stars in consecutive $z_{max}$ intervals for cumulative $e_p$ subsamples are listed and shown in the left panel of Table 1 and Fig. 7, respectively. In Table 1, the planar eccentricity range, the median values of eccentricity, peri- and apo-galactic distances, metallicity, age, the number of RC stars and the radial metallicity gradient results are given for each $z_{max}$ distance interval. Furthermore, if a subsample contains less than 100 RC stars, it is ignored since they lack the statistical sensitivity. 

\begin{figure*}
\centering
\includegraphics[scale=0.38, angle=0]{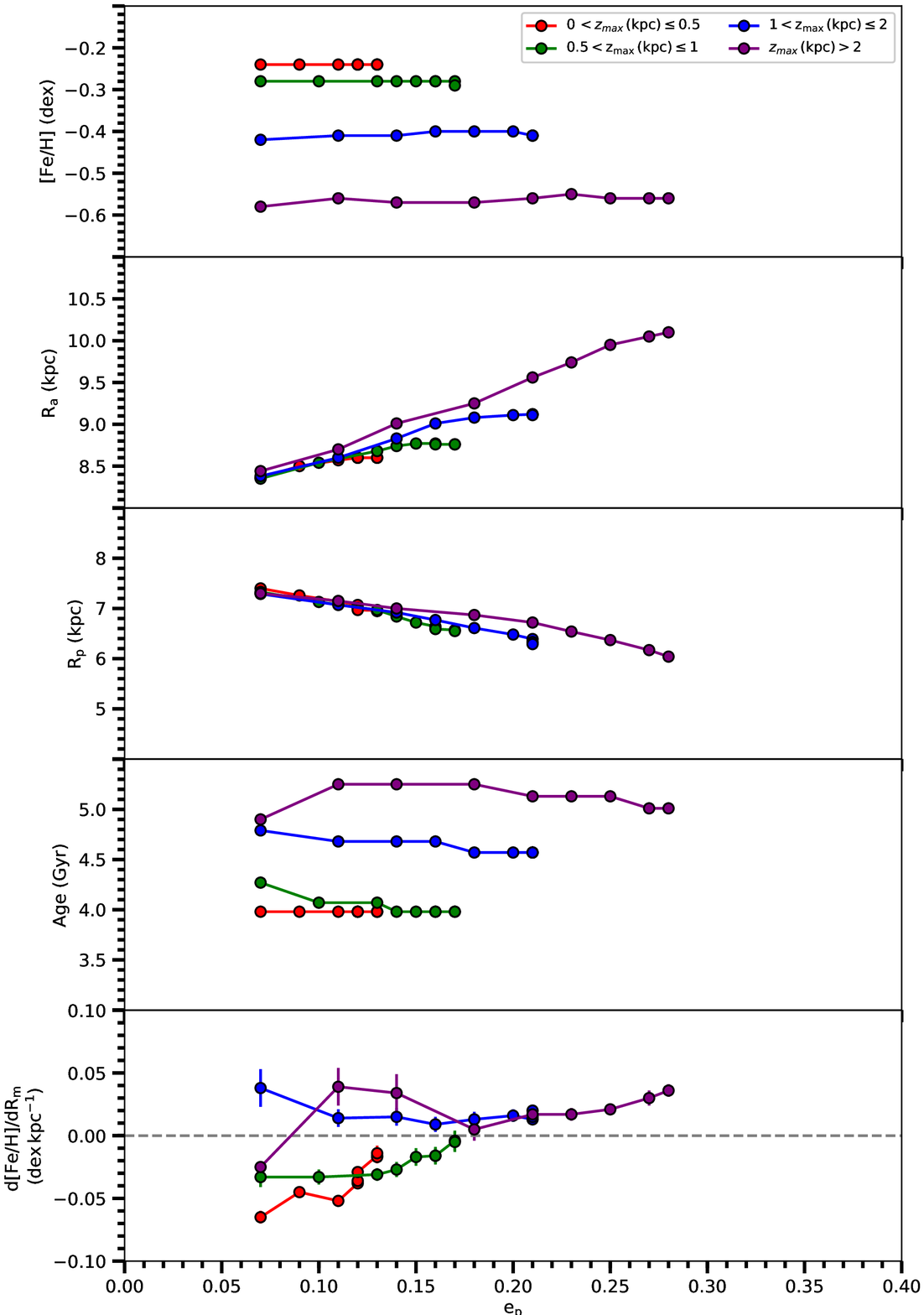}
\includegraphics[scale=0.38, angle=0]{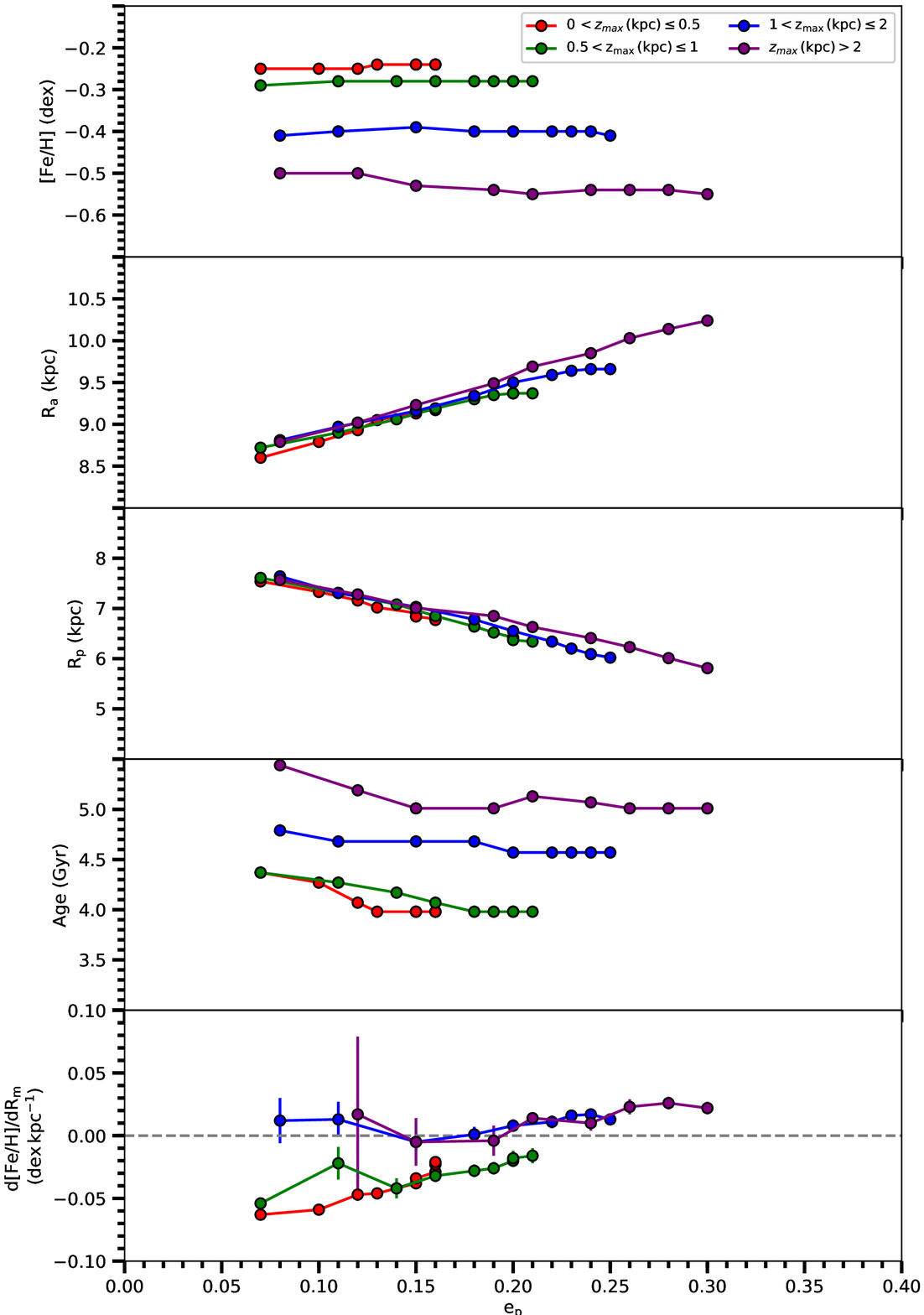}
\caption{Planar eccentricity versus metallicity, apo- and peri-galactic distances, age, and the radial metallicity gradients shown from the top to the bottom panels for the MW (left) and MWBS (right) potential models, respectively. Coloured lines represent the four $z_{max}$ distance intervals, as in Fig. 6.}
\end{figure*} 

\begin{landscape}
\begin{table}%
\centering%
{\scriptsize
\caption{Radial metallicity gradients of the RC stars for four consecutive $z_{max}$ distance intervals and cumulative $e_p$ subsamples. Median values of the planar eccentricity, peri- and apo-galactic distances, metallicity, age, number of stars and radial metallicity gradients calculated from the MW (left panel) and the MWBS (right panel) potential models, respectively.}
\begin{tabular}{cccccccc|cccccccc}%
\hline%
\multicolumn{8}{c|}{MW}&\multicolumn{8}{c}{MWBS}\\%
\hline%
$e_p$&$\tilde{e_p}$&$R_p$&$R_a$&[Fe/H]&$\tau$&$N$&$d${\rm [Fe/H]}$/dR_m$&$e_p$&$\tilde{e_p}$&$R_p$&$R_a$&[Fe/H]&$\tau$&$N$&$d${\rm [Fe/H]}$/dR_m$\\%
 range &  & (kpc) & (kpc) & (dex)& (Gyr) &  & (dex kpc$^{-1}$) &  range &  & (kpc) & (kpc) & (dex)& (Gyr) &  & (dex kpc$^{-1}$) \\%
\hline%
\multicolumn{8}{c|}{$0< z_{max} {\rm (kpc)}\leq 0.5$}&\multicolumn{8}{c}{$0< z_{max}{\rm (kpc)}\leq 0.5$}\\%
\hline%
0.00-0.10&0.07&7.40&8.36&-0.24&3.98& 6007&-0.065$\pm$0.005&0.00-0.10&0.07&7.54&8.60&-0.25&4.37& 3869&-0.063$\pm$0.004\\%
0.00-0.15&0.09&7.26&8.50&-0.24&3.98&10235&-0.045$\pm$0.003&0.00-0.15&0.10&7.33&8.79&-0.25&4.27& 7626&-0.059$\pm$0.004\\%
0.00-0.20&0.11&7.14&8.57&-0.24&3.98&13243&-0.052$\pm$0.004&0.00-0.20&0.12&7.16&8.93&-0.25&4.07&10662&-0.047$\pm$0.005\\%
0.00-0.25&0.12&7.07&8.60&-0.24&3.98&14956&-0.038$\pm$0.003&0.00-0.25&0.13&7.02&9.05&-0.24&3.98&13084&-0.046$\pm$0.003\\%
0.00-0.30&0.12&7.01&8.60&-0.24&3.98&15863&-0.036$\pm$0.002&0.00-0.30&0.15&6.91&9.13&-0.24&3.98&14791&-0.038$\pm$0.003\\%
0.00-0.35&0.12&6.97&8.60&-0.24&3.98&16338&-0.029$\pm$0.002&0.00-0.35&0.15&6.84&9.16&-0.24&3.98&15850&-0.034$\pm$0.003\\%
0.00-0.40&0.13&6.96&8.60&-0.24&3.98&16555&-0.016$\pm$0.005&0.00-0.40&0.16&6.79&9.17&-0.24&3.98&16348&-0.029$\pm$0.002\\%
0.00-0.45&0.13&6.95&8.60&-0.24&3.98&16652&-0.017$\pm$0.005&0.00-0.45&0.16&6.78&9.18&-0.24&3.98&16550&-0.023$\pm$0.002\\%
0.00-0.50&0.13&6.95&8.60&-0.24&3.98&16707&-0.014$\pm$0.006&0.00-0.50&0.16&6.77&9.18&-0.24&3.98&16641&-0.021$\pm$0.003\\%
\hline%
\multicolumn{8}{c|}{$0.5< z_{max}{\rm (kpc)}\leq 1$}&\multicolumn{8}{c}{$0.5< z_{max}{\rm (kpc)}\leq 1$}\\%
\hline%
0.00-0.10&0.07&7.33&8.35&-0.28&4.27& 3750&-0.033$\pm$0.008&0.00-0.10&0.07&7.61&8.72&-0.29&4.37& 2121&-0.054$\pm$0.003\\%
0.00-0.15&0.10&7.13&8.54&-0.28&4.07& 7591&-0.033$\pm$0.006&0.00-0.15&0.10&7.31&8.90&-0.28&4.27& 5072&-0.048$\pm$0.004\\%
0.00-0.20&0.13&6.97&8.68&-0.28&4.07&11254&-0.031$\pm$0.005&0.00-0.20&0.12&7.08&9.06&-0.28&4.17& 8364&-0.042$\pm$0.008\\%
0.00-0.25&0.14&6.84&8.74&-0.28&3.98&13984&-0.027$\pm$0.006&0.00-0.25&0.13&6.85&9.19&-0.28&4.07&11481&-0.032$\pm$0.005\\%
0.00-0.30&0.15&6.72&8.77&-0.28&3.98&15809&-0.017$\pm$0.007&0.00-0.30&0.15&6.64&9.30&-0.28&3.98&14141&-0.028$\pm$0.005\\%
0.00-0.35&0.16&6.64&8.77&-0.28&3.98&16873&-0.016$\pm$0.007&0.00-0.35&0.15&6.52&9.35&-0.28&3.98&15833&-0.026$\pm$0.005\\%
0.00-0.40&0.16&6.59&8.76&-0.28&3.98&17475&-0.016$\pm$0.006&0.00-0.40&0.16&6.42&9.37&-0.28&3.98&16845&-0.020$\pm$0.005\\%
0.00-0.45&0.17&6.57&8.76&-0.28&3.98&17788&-0.004$\pm$0.008&0.00-0.45&0.16&6.37&9.37&-0.28&3.98&17372&-0.018$\pm$0.006\\%
0.00-0.50&0.17&6.55&8.76&-0.29&3.98&17936&-0.005$\pm$0.008&0.00-0.50&0.16&6.34&9.37&-0.28&3.98&17694&-0.016$\pm$0.006\\%
\hline%
\multicolumn{8}{c|}{$1< z_{max}{\rm (kpc)}\leq 2$}&\multicolumn{8}{c}{$1< z_{max}{\rm (kpc)}\leq 2$}\\%
\hline%
0.00-0.10&0.07&7.29&8.38&-0.42&4.79&1085&+0.038$\pm$0.015&0.00-0.10&0.08&7.64&8.81&-0.41&4.79& 591&+0.012$\pm$0.010\\%
0.00-0.15&0.11&7.07&8.60&-0.41&4.68&2449&+0.014$\pm$0.007&0.00-0.15&0.11&7.31&8.97&-0.40&4.68&1574&+0.013$\pm$0.012\\%
0.00-0.20&0.14&6.92&8.83&-0.41&4.68&4152&+0.015$\pm$0.007&0.00-0.20&0.15&7.03&9.16&-0.39&4.68&3007&+0.005$\pm$0.007\\%
0.00-0.25&0.16&6.77&9.01&-0.40&4.68&5783&+0.009$\pm$0.006&0.00-0.25&0.18&6.78&9.34&-0.40&4.68&4529&+0.006$\pm$0.007\\%
0.00-0.30&0.18&6.61&9.08&-0.40&4.57&7150&+0.013$\pm$0.006&0.00-0.30&0.20&6.55&9.50&-0.40&4.57&6013&+0.007$\pm$0.005\\%
0.00-0.35&0.20&6.48&9.11&-0.40&4.57&8078&+0.016$\pm$0.005&0.00-0.35&0.22&6.34&9.59&-0.40&4.57&7196&+0.003$\pm$0.006\\%
0.00-0.40&0.21&6.39&9.12&-0.41&4.57&8649&+0.013$\pm$0.004&0.00-0.40&0.23&6.20&9.64&-0.40&4.57&8032&+0.015$\pm$0.008\\%
0.00-0.45&0.21&6.33&9.12&-0.41&4.57&8985&+0.015$\pm$0.003&0.00-0.45&0.24&6.09&9.66&-0.40&4.57&8547&+0.009$\pm$0.005\\%
0.00-0.50&0.22&6.29&9.12&-0.41&4.57&9194&+0.020$\pm$0.004&0.00-0.50&0.25&6.02&9.66&-0.41&4.57&8852&+0.008$\pm$0.004\\%
\hline%
\multicolumn{8}{c|}{${z_{max} > 2 \, {\rm (kpc)}}$}&\multicolumn{8}{c}{${z_{max} > 2\, {\rm (kpc)}}$}\\%
\hline%
0.00-0.10&0.07&7.30& 8.44&-0.58&4.90& 107&-0.025$\pm$0.000&0.00-0.10&0.08&7.57& 8.79&-0.50&5.44&  58&           --         \\%
0.00-0.15&0.11&7.15& 8.70&-0.56&5.25& 257&+0.039$\pm$0.015&0.00-0.15&0.12&7.28& 9.02&-0.50&5.19& 166&+0.017$\pm$0.062\\%
0.00-0.20&0.14&7.00& 9.01&-0.57&5.25& 453&+0.034$\pm$0.015&0.00-0.20&0.15&7.01& 9.23&-0.53&5.01& 339&-0.005$\pm$0.019\\%
0.00-0.25&0.18&6.87& 9.25&-0.57&5.25& 744&+0.005$\pm$0.009&0.00-0.25&0.19&6.85& 9.49&-0.54&5.01& 573&-0.004$\pm$0.012\\%
0.00-0.30&0.21&6.72& 9.56&-0.56&5.13&1026&+0.017$\pm$0.005&0.00-0.30&0.21&6.63& 9.69&-0.55&5.13& 802&+0.014$\pm$0.004\\%
0.00-0.35&0.23&6.54& 9.74&-0.55&5.13&1253&+0.017$\pm$0.005&0.00-0.35&0.24&6.41& 9.85&-0.54&5.07&1026&+0.010$\pm$0.006\\%
0.00-0.40&0.25&6.37& 9.95&-0.56&5.13&1490&+0.021$\pm$0.004&0.00-0.40&0.26&6.23&10.03&-0.54&5.01&1244&+0.023$\pm$0.006\\%
0.00-0.45&0.27&6.17&10.05&-0.56&5.01&1689&+0.030$\pm$0.006&0.00-0.45&0.28&6.01&10.14&-0.54&5.01&1433&+0.026$\pm$0.005\\%
0.00-0.50&0.28&6.04&10.10&-0.56&5.01&1842&+0.036$\pm$0.005&0.00-0.50&0.30&5.81&10.24&-0.55&5.01&1611&+0.022$\pm$0.005\\%
\hline%
\end{tabular}%
}
\end{table}
\end{landscape}

The median of the planar eccentricities of all RC stars up to $e_p=0.5$ varies between 0.13 and 0.28 when the samples change from distances $0<z_{max}\leq0.5$ kpc to $z_{max}>2$ kpc distances. Median metallicities for each $z_{max}$ interval are found almost constant and become more metal poor as the vertical distance from the Galactic plane is increased. The median metallicities are -0.24, -0.28, -0.41 and -0.56 dex for $0<z_{max}\leq0.5$, $0.5<z_{max}\leq1$, $1<z_{max}\leq 2$ and $z_{max}>2$ kpc, respectively. Similarly, the median ages of the RC stars are 3.98, 3.98, 4.57 and 5.13 Gyrs for the $0<z_{max}\leq 0.5$, $0.5 <z_{max}\leq 1$, $z_{max}\leq 2$ and $z_{max}>2$ kpc distance intervals, respectively. The results show that RC samples are becoming more metal poor and old from the $0<z_{max}\leq 0.5$ to the $z_{max}>2$ kpc distance intervals for the MW potential model. 

The radial metallicity gradients obtained for the MW potential have the highest value for $e_p\leq 0.1$ in the $0<z_{max}\leq 0.5$ kpc distance interval, given by -0.065$\pm$0.005 dex kpc$^{-1}$, and then it decreases to -0.014$\pm$0.006 dex kpc$^{-1}$ in $e_p\leq 0.5$. This implies that the metallicity gradients become positive as the more eccentric orbits added subsequently into the sample. A similar behaviour is found for $0.5< z_{max}\leq1$ kpc distance interval, in which it takes the values -0.033$\pm$0.008 and  -0.005$\pm$0.008 dex kpc$^{-1}$ for $e_p\leq 0.1$ and $e_p\leq 0.5$, respectively. For $1<z_{max}\leq 2$ kpc and $z_{max}>2$ kpc distance intervals, the radial metallicity gradients nearly vanish with tiny values.

\subsubsection{According to MWBS Potential Model}

In order to see the effects of the Galactic bar and the transient spiral arms on the stellar orbits, the mean Galactocentric distances of the RC stars are calculated with the MWBS potential and the sample divided into the aforementioned four $z_{max}$ intervals, accordingly. The results are listed and shown on the right panel of Table 1 and Fig. 7, respectively.

The median values of $e_p$ vary between 0.16 and 0.30 from $0<z_{max}\leq 0.5$ to $z_{max}>2$ kpc distance intervals, respectively. The median metallicities show the same behaviour as for the MW potential, and have values -0.24, -0.28, -0.40 and -0.54 dex for $0<z_{max}\leq 0.5$, $0.5<z_{max}\leq 1$, $z_{max}\leq 2$ and $z_{max}> 2$ kpc, respectively. The median RC ages also progress in the same trend as for the MW potential model, we find 3.98, 3.98, 4.57 and 5.01 Gyrs for $0<z_{max}\leq 0.5$, $0.5<z_{max}\leq1$, $z_{max}\leq2$ and $z_{max}>2$ kpc distance intervals, respectively. The results show that RC samples likewise becoming more metal poor and older from $0<z_{max}\leq0.5$ to $z_{max}>2$ kpc distance intervals, this suggests that the $e_p$, [Fe/H] and $\tau$ parameters calculated for the MWBS potential model show the same trends as seen in the MW potential model.

The radial metallicity gradients obtained under the assumption of the MWBS potential have the steepest value, -0.063 $\pm$ 0.004 dex kpc$^{-1}$, for the $e_p\leq0.1$ subsample in the $0<z_{max}\leq 0.5$ kpc distance interval. The gradients become flatter as more eccentric orbits are added to the subsamples giving, -0.021$\pm$0.003 dex kpc$^{-1}$, for $e_p\leq 0.5$. In $0.5<z_{max}\leq 1$ kpc distance interval, the metallicity gradients tend to become shallower, but still have negative values (about $-0.032$ dex kpc$^{-1}$), like in the $0<z_{max}\leq0.5$ kpc interval, and their values vary between -0.054$\pm$0.003 and -0.016$\pm$0.006 dex kpc$^{-1}$, from $e_p\leq0.1$ to $e_p\leq0.5$, respectively. For the $1< z_{max}\leq 2$ kpc distance interval, the radial metallicity gradients are nearly flat, having positive values closer to zero, except for $e_p\leq 0.1$ where +0.012$\pm$0.010 dex kpc$^{-1}$, but this $e_p$ subsample is affected from the small number of stars, which is almost three times lower than the next $e_p$ subsample. The first subsample, $e_p\leq0.1$, for a distance $z_{max}>2$ kpc is ignored because the number of stars ($N=58$) are statistically low. The radial metallicity gradients tend to remain flat and have positive values.

\subsubsection{Stellar Orbits and Galactic Resonance Regions}
As the number of stars with eccentric orbits increases in the $e_p$ subsamples, the radial metallicity gradients become positive in the $0<z_{max}\leq 0.5$ and $0.5<z_{max}\leq 1$ kpc distance intervals for both the MW and MWBS potential models. It can be deduced that as the planar eccentricities of the disc stars in these distance intervals increase, they move closer towards the Galactic centre, and as a consequence their orbits become perturbed by various phenomena from the inner Galactic regions. In order to investigate these perturbation effects, RC stars in the $z_{max}$ intervals mentioned above are separated into discrete $e_p$ subsamples up to $e_p=0.5$ with $\Delta e_p=0.05$ steps. Output parameters such as the median values of eccentricity, peri- and apo-galactic distances, metallicity, age, the number of RC stars and the radial metallicity gradients are re-calculated for both of the potential models. The results are listed in Table 2.

\begin{table*}[t]
\setlength{\tabcolsep}{4pt}
\centering
{\tiny
\caption{Radial metallicity gradients of the RC stars for four  consecutive $z_{max}$ distances intervals and discrete $e_p$ subsamples. Median values of the planar eccentricity, peri- and apo-galactic distances, metallicity, age, number of stars and radial metallicity gradients calculated from both of the MW (left panel) and MWBS (right panel) potential models.}
\begin{tabular}{cccccccc|cccccccc}%
\hline%
\multicolumn{8}{c|}{MW}&\multicolumn{8}{c}{MWBS}\\%
\hline%
$e_p$&$\tilde{e_p}$&$R_p$&$R_a$&[Fe/H]&$\tau$&$N$&$d${\rm [Fe/H]}$/dR_m$&$e_p$&$\tilde{e_p}$&$R_p$&$R_a$&[Fe/H]&$\tau$&$N$&$d${\rm [Fe/H]}$/dR_m$\\%
 range &  & (kpc) & (kpc) & (dex)& (Gyr) &  & (dex kpc$^{-1}$) & range &  & (kpc) & (kpc) & (dex)& (Gyr) &  & (dex kpc$^{-1}$) \\%
\hline%
\multicolumn{8}{c|}{$0< z_{max} {\rm (kpc)}\leq 0.5$}&\multicolumn{8}{c}{$0< z_{max} {\rm (kpc)}\leq 0.5$}\\%
\hline%
0.00-0.10&0.07&7.40&8.36&-0.24&3.98&6007&-0.065$\pm$0.005&0.00-0.10&0.07&7.54& 8.60&-0.25&4.37&3869&-0.063$\pm$0.004\\%
0.10-0.15&0.12&6.91&8.83&-0.24&3.98&4228&-0.042$\pm$0.005&0.10-0.15&0.12&7.05& 9.02&-0.24&3.98&3757&-0.055$\pm$0.006\\%
0.15-0.20&0.17&6.38&9.01&-0.24&3.98&3008&-0.051$\pm$0.007&0.15-0.20&0.17&6.53& 9.25&-0.24&3.98&3036&-0.042$\pm$0.006\\%
0.20-0.25&0.22&5.76&9.01&-0.24&3.98&1713&-0.025$\pm$0.003&0.20-0.25&0.22&6.01& 9.50&-0.23&3.98&2422&-0.039$\pm$0.004\\%
0.25-0.30&0.27&5.02&8.72&-0.23&3.39& 907&-0.021$\pm$0.007&0.25-0.30&0.27&5.62& 9.87&-0.22&3.63&1707&-0.029$\pm$0.006\\%
0.30-0.35&0.32&4.37&8.50&-0.28&3.98& 475&+0.011$\pm$0.018&0.30-0.35&0.32&5.21&10.21&-0.23&3.89&1059&-0.013$\pm$0.008\\%
0.35-0.40&0.37&3.85&8.43&-0.30&4.07& 217&+0.025$\pm$0.067&0.35-0.40&0.37&4.82&10.55&-0.27&3.80& 498&-0.016$\pm$0.015\\%
0.40-0.45&0.42&3.38&8.39&-0.35&3.24&  97&       --       &0.40-0.45&0.42&4.30&10.75&-0.32&4.07& 202&-0.001$\pm$0.007\\%
0.45-0.50&0.46&2.99&8.33&-0.50&4.68&  55&       --       &0.45-0.50&0.47&3.32& 9.46&-0.36&3.80&  91&     --         \\%
\hline
\multicolumn{8}{c|}{$0.5< z_{max}{\rm (kpc)}\leq 1$}&\multicolumn{8}{c}{$0.5< z_{max}{\rm (kpc)}\leq 1$}\\%
\hline
0.00-0.10&0.07&7.33&8.35&-0.28&4.27&3750&-0.033$\pm$0.008&0.00-0.10&0.07&7.61& 8.72&-0.28&4.37&2121&-0.054$\pm$0.003\\%
0.10-0.15&0.13&6.87&8.82&-0.27&3.98&3841&-0.035$\pm$0.006&0.10-0.15&0.13&7.07& 9.06&-0.28&4.27&2951&-0.025$\pm$0.013\\%
0.15-0.20&0.17&6.46&9.13&-0.28&3.98&3663&-0.025$\pm$0.004&0.15-0.20&0.18&6.56& 9.34&-0.28&3.98&3292&-0.041$\pm$0.008\\%
0.20-0.25&0.22&5.84&9.18&-0.28&3.89&2730&-0.020$\pm$0.007&0.20-0.25&0.22&6.02& 9.52&-0.28&3.98&3117&-0.025$\pm$0.005\\%
0.25-0.30&0.27&5.22&9.13&-0.29&3.80&1825&-0.015$\pm$0.008&0.25-0.30&0.27&5.61& 9.82&-0.28&3.89&2660&-0.025$\pm$0.006\\%
0.30-0.35&0.32&4.53&8.80&-0.31&3.98&1064&-0.005$\pm$0.009&0.30-0.35&0.32&5.21&10.23&-0.28&3.80&1692&-0.020$\pm$0.008\\%
0.35-0.40&0.37&3.95&8.65&-0.38&4.47& 602&+0.015$\pm$0.007&0.35-0.40&0.37&4.77&10.53&-0.30&3.98&1012&-0.005$\pm$0.007\\%
0.40-0.45&0.42&3.36&8.31&-0.46&4.47& 313&+0.070$\pm$0.018&0.40-0.45&0.42&4.06& 9.77&-0.36&4.27& 527&+0.025$\pm$0.014\\%
0.45-0.50&0.47&3.10&8.56&-0.47&4.57& 148&+0.029$\pm$0.005&0.45-0.50&0.47&3.30& 9.26&-0.46&4.57& 322&+0.014$\pm$0.021\\%
\hline
\end{tabular}
}
\end{table*}

According to re-calculations the median values of $e_p$ vary between 0.07 and 0.47 from the $0<e_p\leq0.1$ to the $0.45<e_p\leq0.5$ subsamples, respectively, in the $0<z_{max}\leq 0.5$ and $0.5<z_{max}\leq 1$ kpc distance intervals for both potential models. The peri-galactic distances of the RC stars in both $z_{max}$ intervals for the MW potential model gives 7.40 kpc value for $0<e_p\leq0.1$ subsamples and it decreases down to 2.99 kpc from the Galactic centre for $0.45<e_p\leq0.5$. On the other hand, RC stars are found slightly further away from the Galactic centre for the MWBS potential model, varying between 7.54 and 3.32 kpc for the $0<z_{max}\leq 0.5$ kpc distance interval and between 7.61 and 3.30 kpc for $0.5<z_{max}\leq 1$ kpc distance interval. Median metallicities of the RC sample vary between -0.24 and -0.35 dex for $0<z_{max}\leq 0.5$ kpc in the MW potential model, whereas they vary between -0.25 and -0.36 dex for $0<z_{max}\leq 0.5$ kpc in the MWBS potential model. For $0.5<z_{max}\leq 1$ kpc, median metallicities change between -0.27 and -0.47 dex for the MW potential model and -0.28 and -0.46 dex for the MWBS potential model. When the radial metallicity gradients for both potential models are evaluated, the gradient values are negatively stronger from the $0<z_{max}\leq 0.5$ to the $0.5<z_{max}\leq 1$ kpc distance intervals and also the gradients become shallower with increasing $e_p$ subsamples. The results suggest that RC stars reach out to a distance $R_p\sim3$ kpc from the Galactic centre. When the $e_p$ results obtained from the two potential model are compared, we find that there are more stars in highly eccentric orbits in the MWBS model.

\section{Discussion and Conclusion}

We have investigated the chemo-dynamic structure of the extended Solar neighbourhood calculating the radial metallicity gradients using the mean Galactocentric distances estimated for axisymmetric and non-axisymmetric potential models of 46,818 RC stars from the RAVE DR4 catalogue \citep{Kordopatis13}. This sample is selected by putting constraints on the model atmosphere parameters, namely effective temperature ($4000<T_{eff}\,{\rm (K)}<5400$) and logarithmic surface gravity ($1<\log~g\,{\rm (cgs)}<3$), radial velocity error ($\gamma_{err}\leq10$ km s$^{-1}$), signal to noise ratio ($S/N\geq 40$), and total space velocity error ($S_{err}\leq 20$ km~s$^{-1}$) to select the RC population.   

RC distances are calculated by assuming a mean absolute magnitude of $M_{K_s}=-1.54\pm0.04$ mag \citep{Groenewegen08} in the NIR $K_s$ band, to all objects, due to their standard candle status, using an iterative method which evaluates the colour excess and distances simultaneously, known as the photometric parallax method \citep{Coskunoglu12, Bilir12, Duran13, Karaali14, Plevne15, OnalTas16}. Stellar age of the RC stars are calculated with a method based on the Bayesian statistics \citep{PE04}, while the stellar orbit parameters are calculated by both axisymmetric and a non-axisymmetric potentials, using \textit{MWPotential2014}, in {\it galpy} package \citep{Bovy15}.

Inclusion of the potential functions that represent the Galactic bar and spiral arm potentials into the axisymmetric Milky Way potential altered the stellar orbit parameters, they became elongated and more eccentric. Also, the trends in median planar eccentricities and metallicities show similar characteristics, both have constant values in each $z_{max}$ interval for the MW and MWBS potential models.

Stellar ages become larger with increasing $z_{max}$ distance from the Galactic plane for both of the potential models. The median age is found at about $4\pm1$ Gyr for the entire RC sample. This means that RC stars in RAVE DR4 were born around slightly the same time at different Galactic radii. Thus, the radial metallicity gradients calculated from this data might reflect the chemo-structure of the region which RC stars cover in the Galactic disc. \cite{Girardi16} suggests in his recent review on RC stars that for relatively constant star formation rate the average stellar age for the RC population is somewhere between 1 and 4 Gyrs. This is in accordance with our median RC age for the Solar neighbourhood.

In order to asses the selected effects in metallicity and kinematics in the data, the RAVE DR4 RC sample is divided into three apparent magnitude intervals: $9<I\leq9.75$, $9.75<I<10.4$, and $10.4<I\leq11.3$ mag, which include in the RAVE sample almost the same number of stars (Fig. 8). To assess the bias in metallicities the stellar distribution is plotted for each magnitude interval. We find the mode of the distribution is at -0.30 dex skewed right for each $I$ magnitude interval. Also, the dispersion in metallicity increase as the brightness decreases. A similar assessment is performed for kinematics plotting Toomre diagrams in three $I$ magnitude intervals. In each Toomre diagram, the total space velocity limits are shown with 50 km$^{-1}$ increment in red lines. According to the diagram most of the RC stars lie within $V_{LSR}=150$ km$^{-1}$. Moreover, the total space velocity dispersions vary between 29 and 38 km s$^{-1}$ as the $I$ magnitude intervals increase, but this variation is negligible. These results suggest that RC stars in our sample are analogous in kinematics and metallicity, thus we can conclude that our RAVE DR4 RC sample is unbiased. Similar results are obtained in \citet{Wojno17}'s study that discuss  the selection effects in kinematics and metallicity of all RAVE stars. Finally, this analysis concurs with \citet{Wojno17}'s results.

\begin{figure*}[t]
\centering
\includegraphics[width=16cm, height=12cm]{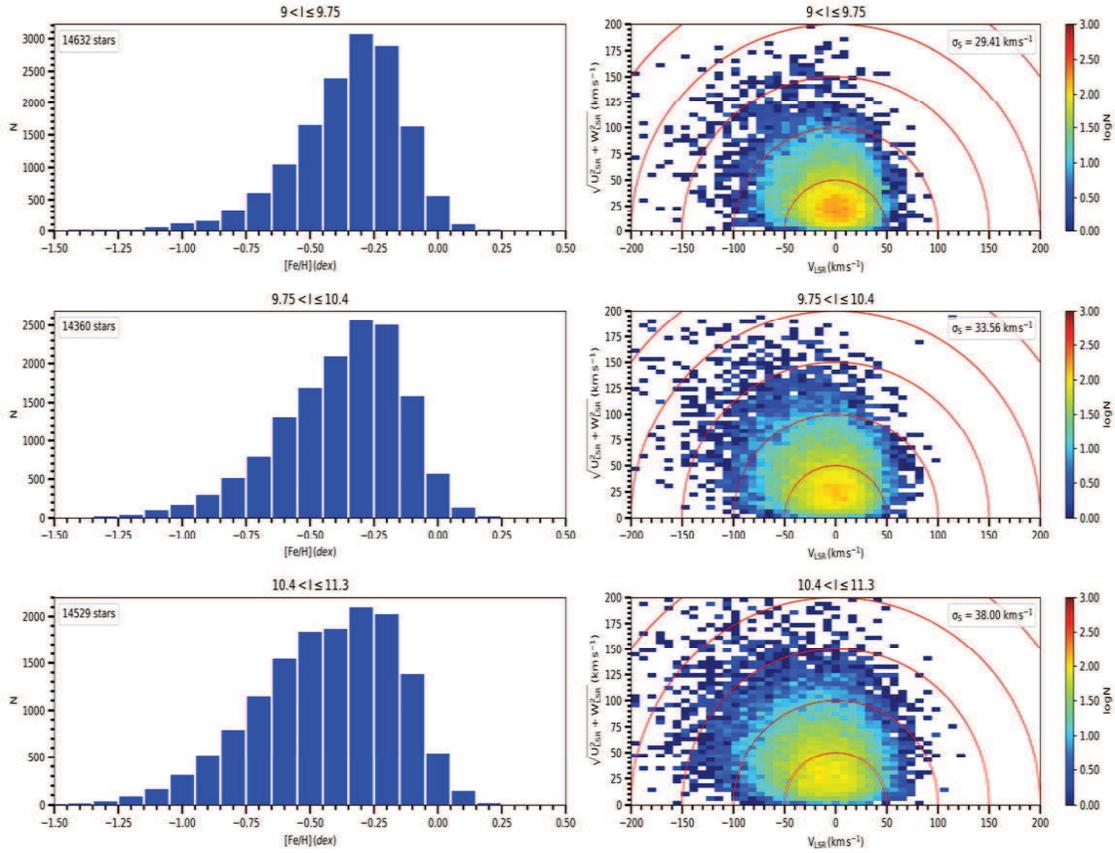}
\caption{Metallicity ({\em left panels}) and Toomre diagrams ({\em right panels}) of RAVE DR4 RC stars in three $I$ apparent magnitude intervals: $9<I\leq9.75$, $9.75<I<10.4$, and $10.4<I\leq11.3$ mag. The colour coding in the Toomre diagrams represent the logarithmic number of stars.} 
\label{Fig08}
\end{figure*}

Tracer objects such as main-sequence, turn-off, red giant branch, and RC stars, cepheid variables, O-B associations, open and globular clusters, planetary nebulae and HII regions, are used in both the radial and vertical metallicity gradients in the literature. These gradients are obtained for three distinct distance indicators, which are the current Galactocentric distance ($R_{gc}$) of stars, the mean Galactocentric distance of the complete orbit of stars ($R_m$) and the guiding radii of the stellar orbits ($R_g$). According to the radial metallicity gradient studies in the literature, the gradient values easily change when the tracer population or the distance indicator is set differently, but their trends do not. This is a challenge to the chemical/cosmic clock approach that serves the logic under the inside-out formation scenario of the Galactic disc. \cite{Jacobson16} used open clusters that lie in the $6 <R_{gc} <12$ kpc distance interval to calculate the radial metallicity gradient as $-0.100\pm0.020$ dex kpc$^{-1}$, while \cite{Genovali14} used cepheid variables that lie in the $5<R_{gc} <19$ kpc distance interval and obtained $-0.021\pm0.029$ dex kpc$^{-1}$. \cite{Chen03} also used Galactic orbit parameters of open clusters that lie within the $6<R_{gc} <15$ kpc distance interval and the obtained gradient is $-0.063\pm0.008$ dex kpc$^{-1}$. Compilation of detailed results of radial and vertical metallicity gradient studies between the years 2000 and 2016 especially from sky surveys are presented in Table 1 of \citet{OnalTas16}. These studies show that the radial metallicity gradients give different results when either the tracer objects are chosen differently, or the radial extents from the Galactic center vary, and the only common property is their negative trend. 

Another perspective lies in the recent studies that use guiding radii as distance indicator for the radial metallicity gradient calculations from RAVE DR4 data. \cite{Plevne15} and \cite{Boeche13} used F-G spectral type main-sequence stars with different sample selection methods and their calculated radial metallicity gradients are $-0.083\pm0.030$ dex kpc$^{-1}$ for $0<z_{max}\leq0.5$ kpc and $-0.065\pm0.003$ dex kpc$^{-1}$ for $z_{max}\leq0.4$ kpc, respectively. Moreover, \cite{Boeche14} calculated metallicity gradients from RC stars and found $-0.054\pm0.004$ dex kpc$^{-1}$ for $z_{max}\leq0.4$ kpc. These results are either caused by the differences in the sample selection or the chosen distance parameter. This shows that even from the same data source the metallicity gradients can give different results. How can  one find the traces of the chemical clocks of the inside-out scenario by not depending on the tracer object.
 
\begin{figure*}[t]
\centering
\includegraphics[scale=0.58, angle=0]{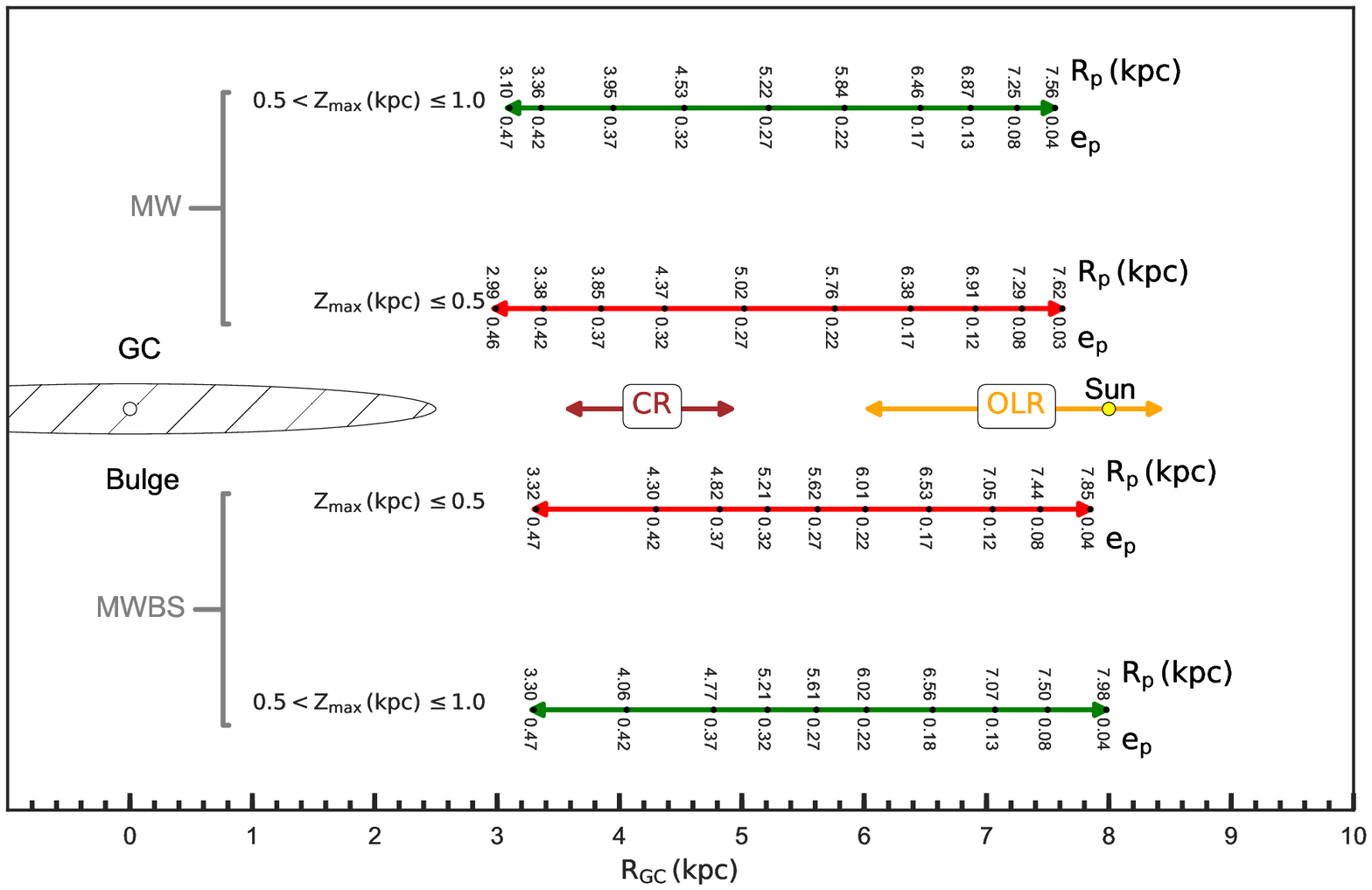}
\caption{Galactic map showing the whereabouts of the Galactic centre, bulge, CR, OLR and the Sun (at $R_{0}=8$ kpc) along with the peri-galactic distances with corresponding planar eccentricities calculated only for $0<z_{max}\leq0.5$ and $0.5<z_{max}\leq1$ kpc distance intervals that show considerable radial metallicity gradients in MW and MWBS potential models. Line colours are the same as in Figs. 6 and 7 for the two $z_{max}$ intervals. The Galactocentric positions of CR and OLR are taken from \cite{Dehnen00}'s study.} 
\end{figure*} 

A new approach to this problem is introduced using the planar eccentricity and maximum vertical distance from the Galactic plane \citep{Plevne15, OnalTas16}. \citet{OnalTas16} is mainly focused on the radial metallicity gradients which are calculated for the mean Galactocentric distances of RC stars selected from the RAVE DR4 catalogue. The median distance of this sample is 1 kpc. The radial metallicity gradient is found as $-0.025\pm0.002$ dex kpc$^{-1}$ for all of the objects in the $0<z_{max}\leq0.5$ kpc distance interval. Then, the planar eccentricity parameter is constrained by cumulatively increasing the $e_p$ subsamples. The steepest radial metallicity gradient value is $-0.089\pm0.002$ dex kpc$^{-1}$ for the $e_p\leq0.05$ subsample. This result is in good agreement with the metallicity gradient of F-G spectral type main-sequence sample of \cite{Plevne15}, constrained by $e_p\leq0.1$ and $z_{max}\leq825$ pc. They found the gradient in the $0<z_{max}\leq0.5$ kpc distance interval as $-0.083\pm0.030$ dex kpc$^{-1}$, as well. Both samples are representing both the Galactic thin disc and open star clusters. These metallicity gradient studies obtained with RAVE data are also in good agreement with \cite{Wu09}'s study in which they determined the open cluster radial metallicity gradient in $R_m$, for stars that lie in the $6<R_{gc}<20$ kpc distance interval close to the Galactic plane and find, $-0.082\pm0.014$ dex kpc$^{-1}$. These findings suggest that the radial metallicity gradients are a function of $e_p$ and $z_{max}$. These dependencies can be determined by investigating the effects of phenomena in the Galactic disc.

This study introduces the planar eccentricity limits that coincide with the Galactocentric radii where the positions of the well-known resonance regions. Radial metallicity gradients show a general trend of flattening with increasing planar eccentricity and also their values become more positive with increasing vertical distance from the Galactic plane. This trend has been found for both of the potential models. We noticed that once we eliminated the RC stars with more eccentric orbits, the radial metallicity gradients become steeper. We also put a schematic diagram for RC stars which show their orbital extensions towards the inner Galaxy. Fig. 9 helps to spot whether the sample stars are inside a resonance region, i.e. CR or OLR, or not, from the interrelation between the planar eccentricity and peri-galactic distance parameters in $0<z_{max}\leq0.5$ and $0.5<z_{max}\leq 1$ kpc intervals for the MW and MWBS potential models. The radial coverage of the CR and OLR regions are taken from \citet{Dehnen00}. According to the schematic diagram, RC stars under the influence of an axisymmetrical potential reach down to 3 kpc distance from the Galactic centre, whereas for the MWBS potential model the peri-galactic distances increase by $\sim 0.3$ kpc ($R_p=3.3$ kpc). This is another demonstration of what happens to stellar orbits when major perturbers are considered in potential calculations. The schematic diagram in Fig. 9 shows that the limit of OLR is where the RC stars have planar eccentricities between $0<e_p<0.2$ for both $z_{max}$ intervals for the MW potential model, whereas the planar eccentricity range is slightly larger, $0<e_p<0.22$, for both $z_{max}$ distance intervals for the MWBS potential model. Meanwhile, there is almost a 1 kpc gap between CR and OLR regions and this is reflected on planar eccentricities of the MW potential as $0.20<e_p<0.27$ for the $0<z_{max}\leq0.5$ kpc and $0.5<z_{max}\leq 1$ kpc distance intervals. Results from the MWBS potential seem to be more compact and vary within $0.22<e_p<0.36$. Based on Fig. 9, stellar orbits calculated from the MW potential model reach beyond the CR region, meanwhile stellar orbits of the MWBS potential model only reach into the vicinity of the CR region. The radial metallicity gradients calculated from the discrete $e_p$ subsamples are becoming shallow as the eccentricity ranges increase. In Fig. 9, the RC subsamples with eccentric orbits reach out to 3 kpc distance towards the Galactic center. Thus, once the samples that might originate from slightly inner Galactic radii are excluded, then the radial metallicity gradients become more authentic for the Solar neighbourhood. We claim that this apparent change in radial metallicity gradients in the thin disc is a result of the RC stars perturbed from the existing resonance regions, the largest radial metallicity gradients are obtained where the outer Lindblad resonance region is effective.

Ultimately, this study can be improved by combining the model atmosphere parameters and radial velocities coming from new HRS observations and trigonometric parallaxes and proper motions coming from the {\em Gaia} astrometry satellite. This will provide more reliable inputs from kinematic and dynamic orbit analysis and give more realistic results about the perturbation effects in the inner Galaxy and their reflection in the Solar neighbourhood.

\section{Acknowledgments}
The authors are grateful to the anonymous referee for his/her considerable contributions to improve the paper. This study has been supported in part by the Scientific and Technological Research Council (T\"UB\.ITAK) 114F347 and the Research Fund of the University of Istanbul, Project Number: 24292. This study is a part of the PhD thesis of \"Ozgecan \"Onal Ta\c s. This research has made use of NASA's (National Aeronautics and Space Administration) Astrophysics Data System and the SIMBAD Astronomical Database, operated at CDS, Strasbourg, France and NASA/IPAC Infrared Science Archive, which is operated by the Jet Propulsion Laboratory, California Institute of Technology, under contract with the National Aeronautics and Space Administration. Also, the authors are grateful to Prof. Dr. Salih Karaali and Dr. Kai O. Schwenzer for their contributions on revising the paper.

\end{document}